\documentclass[twocolumn,english,aps,pra,showpacs]{revtex4}

\usepackage{color,rotating,graphicx,subfigure}
\usepackage{amsmath,amssymb,amscd}

\usepackage[latin1]{inputenc}
\usepackage[english]{babel}

\usepackage{dsfont}
\usepackage{textcomp}

\renewcommand{\Re}{{\rm Re}}
\renewcommand{\Im}{{\rm Im}}

\newcommand{\ri}{{\rm i}}
\newcommand{\re}{{\rm e}}
\newcommand{\rd}{{\rm d}}

\newcommand{\rc}{{\rm c}}

\newcommand{\rs}{{\rm s}}
\newcommand{\rp}{{\rm p}}  
\newcommand{\ro}{{\rm o}}

\begin{document}
\title{Qubit entanglement across Epsilon near Zero Media}

\author{S.-A. Biehs} 

\affiliation{Institut f\"{u}r Physik, Carl von Ossietzky Universit\"{a}t, D-26111 Oldenburg, Germany}

\author{G. S. Agarwal}

\affiliation{ Institute for Quantum Science and Engineering and Department of Biological and 
Agricultural Engineering, Texas A \& M University, College Station, Texas 77845, USA}

\email{s.age.biehs@uni-oldenburg.de}

\begin{abstract}
Currently epsilon near zero materials (ENZ) have become important for controlling the propagation of light and enhancing by several orders of  magnitude the Kerr and  other  nonlinearities. Given this advance it is important to examine the quantum electrodynamic processes and information tasks near ENZ materials. We study the entanglement between two two-level systems near ENZ materials and compare our results with the case where the ENZ material is replaced by a metal.
It is shown that with ENZ materials substantial entanglement can be achieved over larger distances than for metal films. We
show that this entanglement over large distances is due to the fact that one can not only have large emission rates but also large energy transmission rates at the epsilon-near-zero wavelength. This establishes superiority of ENZ  materials for studying processes specifically important for quantum information tasks.
\end{abstract}

\maketitle



\section{Introduction}

A large number of problems in physics and chemistry require very significant 
dipole-dipole interaction which includes fundamental interactions such as
 van der Waals forces and vacuum friction~\cite{MilonniBook,Volokitin2007}, 
F\"{o}rster (radiative) energy transfer (FRET)~\cite{Forster,DungEtAl2002}, radiative heat 
transfer~\cite{Volokitin2007,BiehsAgarwal2013b}, quantum information protocols 
like the realization of CNOT gates~\cite{Nielsen,Bouchoule2002,Isenhower2010}, 
pairwise excitation of atoms~\cite{VaradaGSA1992,HaakhCano2015,HettichEtAl2002}, 
and Rydberg blockade~\cite{SaffmanEtAl,GilletEtAl}. In the last decades
numerous plasmonic and metamaterial platforms have been developed to enhance
the dipole-dipole interaction significantly. For example for FRET it could be
shown theoretically and experimentally that when two atoms placed in the vicinity 
of 2D or 2D-like plasmonic structures as graphene 
sheets~\cite{Velizhanin2012, AgarwalBiehs2013,KaranikolasEtAl2016} 
and metal films~\cite{BiehsAgarwal2013,Poudel2015,LiEtAl2015,CanoEtAl2010,BouchetEtAl2016} 
can persist over long distances due to the plasmon assisted energy transfer. 
More astonishingly is that one can even find a significant energy 
transfer across metal films~\cite{AndrewBarnes2004} due to the interaction
with the coupled surface plasmons which can be highly improved by replacing the
metal film by a hyperbolic meta-material~\cite{BiehsEtAl2016}. This effect of 
a long-range energy transfer across a hyperbolic meta-material can be regarded
as one form of the so-called super-Coulombic atom-atom interaction~\cite{CortesJacob2017}.

Currently epsilon near zero materials (ENZ) have become important for controlling the 
propagation of light~\cite{Engheta2016} and enhancing by several orders of magnitude the Kerr and other 
nonlinearities~\cite{Boyd2016}. Given this advance it is important to examine the quantum electrodynamic 
processes and information tasks near ENZ materials. The ENZ media can also be used to 
increase tunneling electromagnetic energy through subwavelength channels~\cite{Silverinha2006}, to allow for
phase-pattern tailoring~\cite{Alu2007}. Further application for control of the emission of quantum 
emitters in open ENZ cavities has been discussed~\cite{Engheta2016}. In this work we will show that with 
multilayer hyperbolic meta-materials substantial entanglement in the visible regime can 
be produced over larger distances than with metallic films. Especially, at the ENZ wavelength 
one can have large emission rates, energy transfer rates and entanglement when using hyperbolic 
metamaterials, which have already been shown to be very advantageous for energy transfer and heat
transfer~\cite{LangEtAl2014,LiuEtAl2015,BiehsEtAlPRL2015,BiehsEtAl2016,MessinaEtAl2016}. 
In contrast, for metals we find that using the ENZ wavelength is 
not advantageous for long distance entanglement, so that the anisotropic character
of hyperbolic materials is the driving factor for the observed effect.

\begin{figure}[hbt]
  \begin{center}
    \includegraphics[width = 0.35\textwidth]{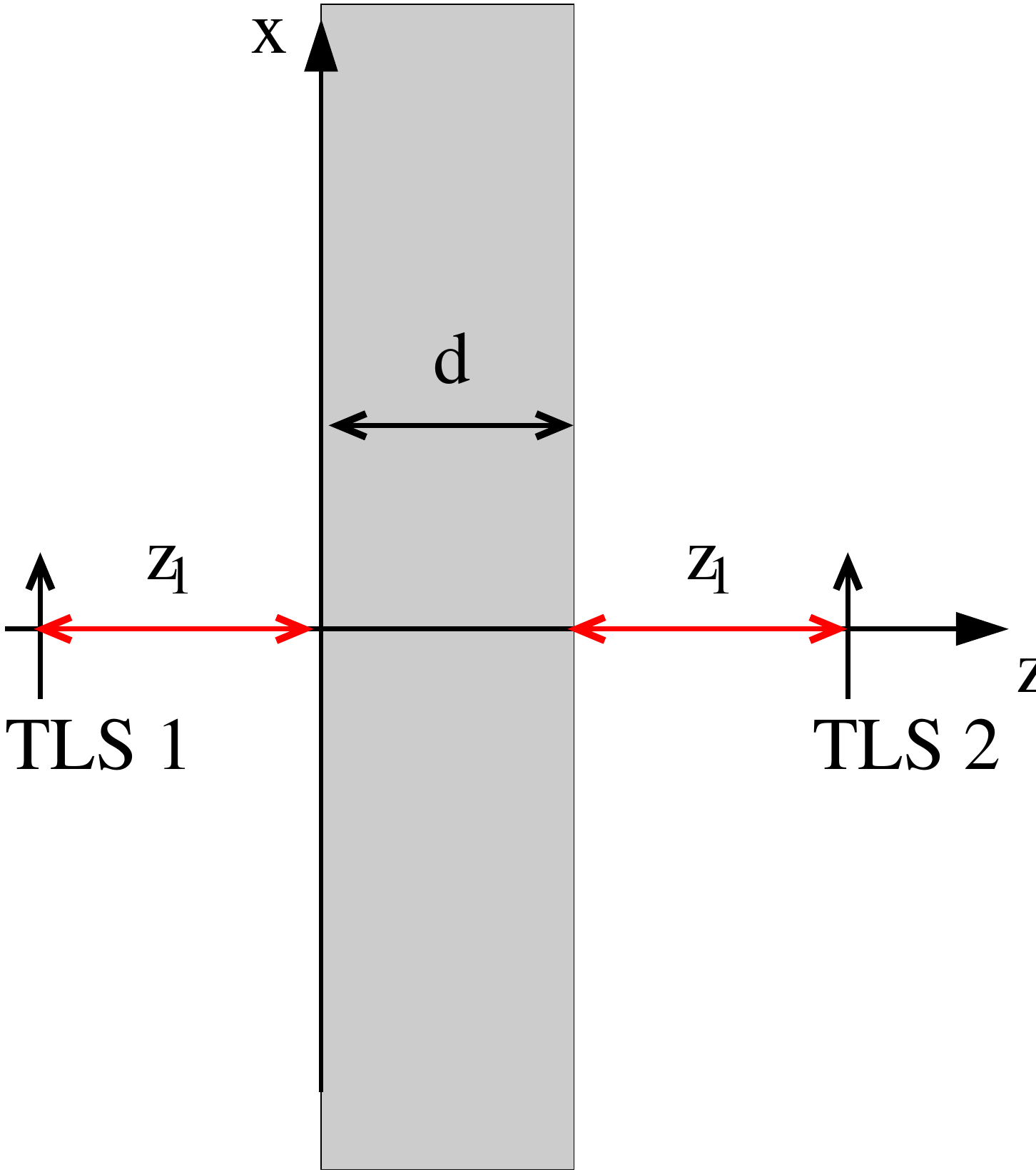}
  \end{center}
  \caption{Sketch of the considered configuration.\label{Fig:SketchGeometry}}
\end{figure}

The paper is organized as follows: In Sec.~II we introduce the model and
the general expressions needed to determine the concurrence function of
two coupled TLS in the presence of a plasmonic environment which is 
described by the Green's function given in Sec.~III. In Sec.~IV we
compare the degree of entanglement between the TLS separated which can
be achieved with a thin silver film with that of mulilayer hyperbolic 
meta-material. The conclusions of our study are given in Sec.~V.

%
%
\section{Calculation of entanglement measure}

The quantum entanglement arises from the radiative coupling between the two dipoles. 
Initially the atom A is excited and the atom B is in ground state. Thus to start with there is 
no entanglement between A and B atoms. When the atom A emits photon then this photon can be 
absorbed by the atom B leading to its excitation. This process can go on. Thus the quantum 
entanglement is produced by the dynamical evolution of the system of atoms. The dynamical 
evolution is most conveniently described in the master equation framework. The master equation 
is obtained by eliminating the radiative degrees of freedom and depends on the plasmonic or 
hyperbolic environment in which the atoms are located~\cite{Agarwal1975,AgarwalBook}. The density matrix of the 
two atoms is given by the environment dependent master equation ($i,j = 1,2$)
\begin{equation}
\begin{split}
  \frac{\partial \rho}{\partial t} &= - \ri \omega_0 \sum_j [S_j^z, \rho] - \ri \sum_{i,j} \Omega_{ij} [S_i^+ S_j^-, \rho] \\
                                   &\quad - \sum_{i,j} \gamma_{ij} [S_i^+ S_j^- \rho - 2 S_j^- \rho S_i^+ + \rho S_i^+ S_j^-]
\end{split}
\end{equation}
where $\omega_0$ is the transition frequency of the two TLS, $S^+_j = |e_j\rangle \langle g_j|$ and $S^-_j = |g_j\rangle \langle e_j |$ are the atomic ladder operators and $S_j^z = \frac{1}{2}(|e_j\rangle \langle e_j| - |g_j\rangle \langle g_j|)$. The coupling of the two TLS via its environment which functions as a reservoir is described by
\begin{align}
  \gamma_{ij} &:=   \frac{1}{\epsilon_0}\Im\biggl( \frac{\omega_0^2}{c^2} \frac{1}{\hbar } \mathbf{p}_i\cdot\mathds{G}(\mathbf{r}_i,\mathbf{r}_j, \omega_0)\cdot\mathbf{p}_j^* \biggr), \\
  \Omega_{ij} &:= - \frac{1}{\epsilon_0}\Re\biggl( \frac{\omega_0^2}{c^2} \frac{1}{\hbar} \mathbf{p}_i\cdot\mathds{G}(\mathbf{r}_i,\mathbf{r}_j, \omega_0)\cdot\mathbf{p}_j^* \biggr)
\end{align} 
introducing the dyadic Green's function $\mathds{G}$ which determines the entire dynamics
of our system. Typically, it can be written as a sum 
$\mathds{G} = \mathds{G}^{(0)} + \mathds{G}^{(s)}$ of the vacuum and a scattering
part which takes the presence of the plasmonic or hyperbolic environment into account. 
Here, $\gamma_{11}$, $\gamma_{22}$ and $\Omega_{11}$, $\Omega_{22}$ are the single
atom emission rates and level shifts of TLS 1 and 2 in the presence of the plasmonic 
environment, whereas $\gamma_{12}, \gamma_{21}, \Omega_{21}$ and $\Omega_{12}$ are
the corresponding collective damping rates and level shifts.
By writing the dipole moment of the two TLS as
\begin{equation}
  \mathbf{p}_i := p \mathbf{e}_i,
\end{equation}
where $\mathbf{e}_i$ is the general complex valued unit vector pointing in the direction of the dipole moment of TLS $i$, and introducing the free space emission rate of a single TLS~\cite{AgarwalBook}
\begin{equation}
  \gamma_0 := \frac{2 |p|^2 \omega_0^3}{3 c^3 \hbar} \frac{1}{4 \pi \epsilon_0} 
\end{equation}
we can express $\gamma_{ij}$ and $\Omega_{ij}$ as
\begin{align}
  \gamma_{ij} &:=   6 \pi \gamma_0 \frac{c}{\omega_0} \Im\biggl(\mathbf{e}_i\cdot\mathds{G}(\mathbf{r}_i,\mathbf{r}_j, \omega_0)\cdot\mathbf{e}_j^* \biggr), \\
  \Omega_{ij} &:= - 6 \pi \gamma_0 \frac{c}{\omega_0} \Re\biggl(\mathbf{e}_i\cdot\mathds{G}(\mathbf{r}_i,\mathbf{r}_j, \omega_0)\cdot\mathbf{e}_j^* \biggr).
\end{align} 
If we consider a symmetric configuration as for example depicted in Fig.~\ref{Fig:SketchGeometry}, we further have
\begin{align}
  \gamma_{12} = \gamma_{21} =: \gamma_\rc, \\
  \gamma_{11} = \gamma_{22} =: \gamma_\rs, \\
  \Omega_{12} = \Omega_{21} =: \Omega_\rc, \\
  \Omega_{11} = \Omega_{22} =: \Omega_\rs 
\end{align}
where the indices stand for 'single' and 'collective'.
In this case we find the dynamical equations
\begin{align}
  \dot{\rho}_{ee}  &= - 4 \gamma_\rs \rho_{ee}, \\
  \dot{\rho}_{eg}  &= -2 \rho_{eg} [\gamma_\rs + \ri (\omega_0 + \Omega_\rs) ], \\
  \dot{\rho}_{ss}  &= -2 (\rho_{ss} - \rho_{ee}) (\gamma_\rs + \gamma_\rc), \\
  \dot{\rho}_{aa}  &= -2 (\rho_{aa} - \rho_{ee}) (\gamma_\rs - \gamma_\rc), \\
  \dot{\rho}_{as}  &= -2 \rho_{as} (\gamma_\rs - \ri \Omega_\rc)
\end{align}
for the states $|e\rangle = |e_1 e_2\rangle$, $|g\rangle = | g_1 g_2 \rangle$, $|s\rangle = (| e_1 g_2\rangle + |g_1 e_2 \rangle ) / \sqrt{2}$ and $|a\rangle = (| e_1 g_2\rangle - |g_1 e_2 \rangle ) / \sqrt{2}$.
These equations correspond to Eqs.~(15.27) in Ref.~\cite{AgarwalBook}. Accordingly, the solutions of the
dynamical equations are
\begin{align}
   \rho_{ee}(t) &= \rho_{ee}(0) \re^{- 4 \gamma_\rs t}, \\
   \rho_{eg}(t) &= \rho_{eg}(0) \re^{-2 [\gamma_\rs + \ri (\omega_0 - \Omega_\rs )] t}, \\
   \rho_{as}(t) &= \rho_{as}(0) \re^{-2 (\gamma_\rs - \ri \Omega_\rc) t}, \\
   \rho_{aa}(t) &= \rho_{aa}(0) \re^{-2(\gamma_\rs - \gamma_\rc) t} \nonumber \\
                &\quad  - \frac{\gamma_\rs - \gamma_\rc}{\gamma_\rs + \gamma_\rc} \rho_{ee}(0) \biggl( \re^{- 4 \gamma_\rs t} - \re^{-2 (\gamma_\rs - \gamma_\rc) t} \biggr), \\
   \rho_{ss}(t) &=  \rho_{ss}(0) \re^{-2(\gamma_\rs + \gamma_\rc) t} \nonumber \\ 
                &\quad- \frac{\gamma_\rs + \gamma_\rc}{\gamma_\rs - \gamma_\rc} \rho_{ee}(0) \biggl( \re^{- 4 \gamma_\rs t} - \re^{-2 (\gamma_\rs + \gamma_\rc) t} \biggr).
\end{align}
From these equations the dynamics of the system of TLS coupled by an arbitrary environment
can be studied. Here we are interested in the entanglement which is measured by the concurrence 
function $C(t)$ introduced by Wootters~\cite{Wootters1998}. This function is defined by
$C = \max(0, \sqrt{\lambda_1} - \sqrt{\lambda_2} - \sqrt{\lambda_3} - \sqrt{\lambda_4})$
where the $\lambda_i$ ($i = 1,\ldots, 4$) are the eigenvalues of the matrix $\rho \tilde{\rho}$;
$\tilde{\rho}$ can be defined by means of the Pauli matrix $\sigma_y$ by 
$\tilde{\rho} = \sigma_y \otimes \sigma_y \rho^* \sigma_y \otimes \sigma_y$. The concurrence
functions has values between $[0,1]$ giving $0$ for unentangled states and $1$ for maximally 
entangled states. When starting at $t = 0$ with the initially unentangled 
state $|e_1,g_2 \rangle$ then the concurrence function $C(t)$ is given by~\cite{TanasFicek2004,AgarwalBook} 
\begin{equation}
  C(t) = \re^{-2 \gamma_\rs t} \sqrt{\sinh^2(2 \gamma_\rc t) + \sin^2(2 \Omega_\rc t)}.
\label{Eq:Concurrence}
\end{equation}
It can be seen that $C(t = 0) = 0$ as expected, but for times $t > 0$ it can have values
larger than zero which means that due to the coupling of the two TLS via the environment
an entanglement of the states of the two TLS is produced.  

%
%

\section{Dyadic Green's function}
 
The goal is now to study the entanglement measured by the concurrence function 
for the two TLS when they are coupled by 
a thin film as depicted in Fig.~\ref{Fig:SketchGeometry}. To this end, it is necessary to determine
$\gamma_\rc$, $\gamma_\rs$ and $\Omega_\rc$ which means that we have to determine the
corresponding Green's function for that configuration. For an initially excited TLS 
at $\mathbf{r}_1  = (0,0,z_1)^t$  with $z_1 < 0$ and a second TLS which is initially in 
the ground state at $\mathbf{r}_2  = (0,0,z_2 = d + |z_1|)^t$ the Green's function 
in Weyl's representation is given by
\begin{equation}
  \mathds{G}(\mathbf{r_1,r_2}) =  \int \!\!\frac{\rd^2 \kappa}{(2 \pi)^2} \, \mathds{G}(\boldsymbol{\kappa},z)
\end{equation}
where $\boldsymbol{\kappa} = (k_x,k_y)^t$ and 
\begin{equation}
   \mathds{G} (\boldsymbol{\kappa},z) = \frac{\ri \re^{\ri k_{z,{\rm vac}} (d + 2|z_1|)}}{2 k_{z,{\rm vac}}} \sum_{i = \rs,\rp} t_i \mathbf{a}_i^+(k_0) \otimes \mathbf{a}_i^+(k_0)
\end{equation}
introducing the vacuum wavevector in z-direction $k_{z,{\rm vac}} = \sqrt{k_0^2 - \kappa^2}$ 
and $k_0 = \omega_0/c$. Here $t_\rs$ and $t_\rp$ are the amplitude transmission coefficients 
and $\mathbf{a}_{\rs,\rp}^+$ are the polarization vectors defined by 
\begin{align}
  \mathbf{a}_\rs^+ (k_0) &= \frac{1}{\kappa} \begin{pmatrix} k_y \\ - k_x \\ 0 \end{pmatrix}, \\ 
  \mathbf{a}_\rp^+(k_0) &= \frac{1}{\kappa k_0}\begin{pmatrix} - k_x k_{z,{\rm vac}} \\ - k_y k_{z,{\rm vac}} \\ \kappa^2 \end{pmatrix}.
\end{align}
From this expression we can determine $\gamma_\rc$ and $\Omega_\rc$. 
On the other hand, if we want to determine $\gamma_\rs$ we have to determine the 
the Green's function $\mathds{G}(\mathbf{r_1,r_1})$ evaluated solely at the position of the
TLS which is initially in the excited state. In this case, we have
\begin{equation}
  \mathds{G}(\mathbf{r}_1,\mathbf{r}_1) =  \int \!\!\frac{\rd^2 \kappa}{(2 \pi)^2} \, \mathds{G}^{\rm single}(\boldsymbol{\kappa},z)
\end{equation}
with 
\begin{equation}
\begin{split}
   \mathds{G}^{\rm single} (\boldsymbol{\kappa},z_1) &= \frac{\ri}{2 k_{z,{\rm vac}}} \sum_{i = \rs,\rp} \biggl[ \mathbf{a}_i^+(k_0) \otimes \mathbf{a}_i^+(k_0) \\  &\quad + \re^{2 \ri k_{z,{\rm vac}} |z_1|} r_i \mathbf{a}_i^+(k_0) \otimes \mathbf{a}_i^-(k_0) \biggr],
\end{split}
\end{equation}
where $r_\rs$ and $r_\rp$ are the amplitude reflection coefficients.

In order to see how the entanglement is affected by a plasmonic structure like 
a simple metal film or a hyperbolic meta-material we only need the appropriate
expressions for the transmission and reflection coefficients. These are well 
known and for a in general uni-axial material with the optical axis oriented
along the surface normal the transmission and reflection coefficients are given by
\begin{align}
  t_\rs &= \frac{4 k_{z,\ro} k_{z,{\rm vac}} \re^{\ri (k_{z,\ro} - k_{z,{\rm vac}}) d}}{(k_{z,\ro} + k_{z,{\rm vac}})^2 - (k_{z,\ro} - k_{z,{\rm vac}})^2 \re^{2 \ri k_{z,\ro} d}} \\
  t_\rp &= \frac{4 \epsilon_\parallel k_{z,\re} k_{z,{\rm vac}} \re^{\ri (k_{z,\re} - k_{z,{\rm vac}}) d}}{(k_{z,\re} + \epsilon_\parallel k_{z,{\rm vac}})^2 - (k_{z,\re} - \epsilon_\parallel k_{z,{\rm vac}})^2 \re^{2 \ri k_{z,\re} d}},
\end{align}
and 
\begin{align}
  r_\rs &= R_\rs \frac{1 - \re^{2 \ri k_{z,\ro} d}}{1 - R_\rs^2 \re^{2 \ri k_{z,\ro} d}}, \\
  r_\rp &= R_\rp \frac{1 - \re^{2 \ri k_{z,\re} d}}{1 - R_\rp^2 \re^{2 \ri k_{z,\re} d}}, 
\end{align}
where we have introduced the reflection coefficients of a single interface
\begin{align}
  R_\rs &= \frac{k_{z,{\rm vac}} - k_{z,\ro}}{k_{z,{\rm vac}} + k_{z,\ro}}, \\
  R_\rp &= \frac{k_{z,{\rm vac}} \epsilon_\perp  - k_{z,\re}}{k_{z,{\rm vac}} \epsilon_\perp + k_{z,\re}}
\end{align}
and the z components of the wavevector for the ordinary and extraordinary modes
\begin{align}
  k_{z,\ro} &= \sqrt{k_0^2 \epsilon_\perp - \kappa^2}, \\ 
  k_{z,\re} &= \sqrt{k_0^2 \epsilon_\perp - \kappa^2 \frac{\epsilon_\perp}{\epsilon_\parallel}}.
\end{align}
The permittivities $\epsilon_\perp$ and $\epsilon_\parallel$ are the permittivities
perpendicular and parallel to the optical axis which is here the z axis, i.e. the
optical axis is along the surface normal.

%
%
\section{Metal vs HMM films}

In the following we will consider a single silver film described by the Drude model
\begin{equation}
  \epsilon_{\rm Ag} = \epsilon_\parallel = \epsilon_\perp = \epsilon_\infty - \frac{\omega_\rp^2}{\omega(\omega + \ri \tau^{-1})}.
\label{Eq:Drude}
\end{equation}
The parameters from~\cite{ChristyJohnson,Soennichsen} are $\epsilon_\infty = 3.7$, $\omega_\rp = 1.4\cdot10^{16}\,{\rm rad}/{\rm s}$,$\tau = 4\cdot10^{-14}\,{\rm s}$. As in Ref.~\cite{BiehsEtAl2016} we use a much smaller relaxation time of $\tau = 0.45\cdot10^{-14}\,{\rm s}$ which accounts for the increased collission frequency found in thin metal films~\cite{LangEtAl2013}. The surface plasmon resonance wavelength is in this case given by $\lambda_{\rm SP} = 291\,{\rm nm}$ and the ENZ wavelength is $\lambda_{\rm ENZ} = 259\,{\rm nm}$.

\begin{figure*}[hbt]
  \begin{center}
 \includegraphics[width = 0.45\textwidth]{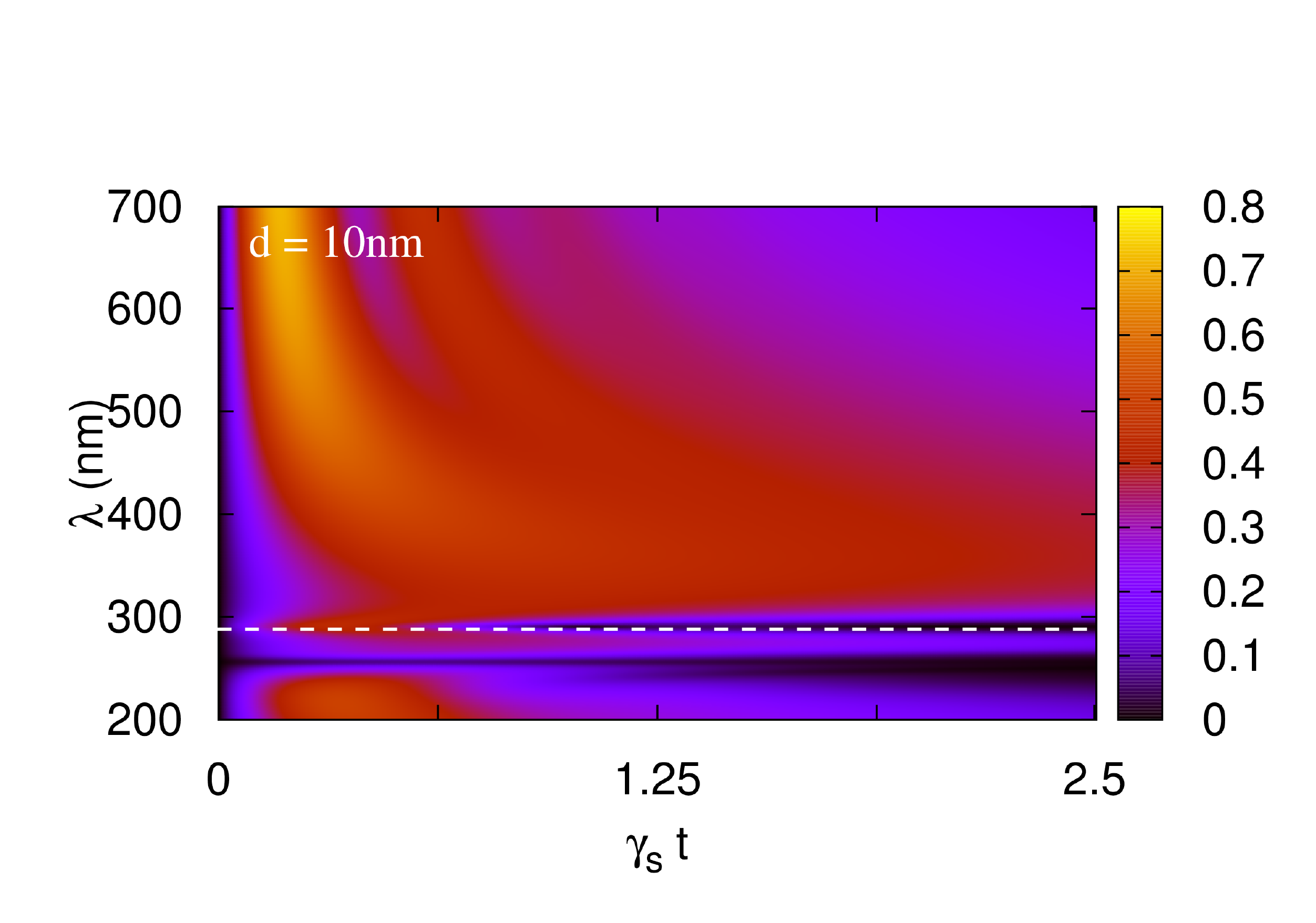} 
 \includegraphics[width = 0.45\textwidth]{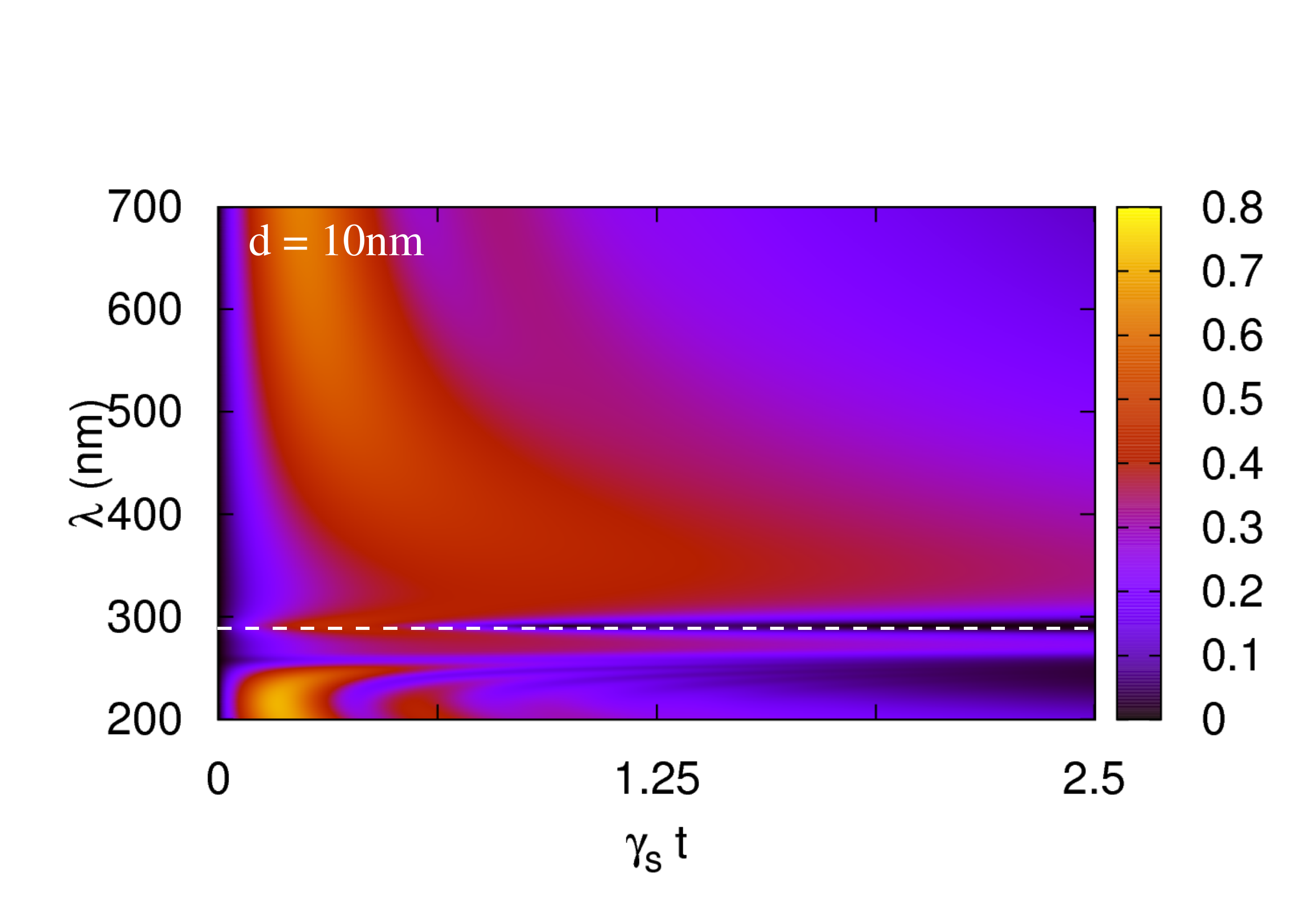} \\
 \includegraphics[width = 0.45\textwidth]{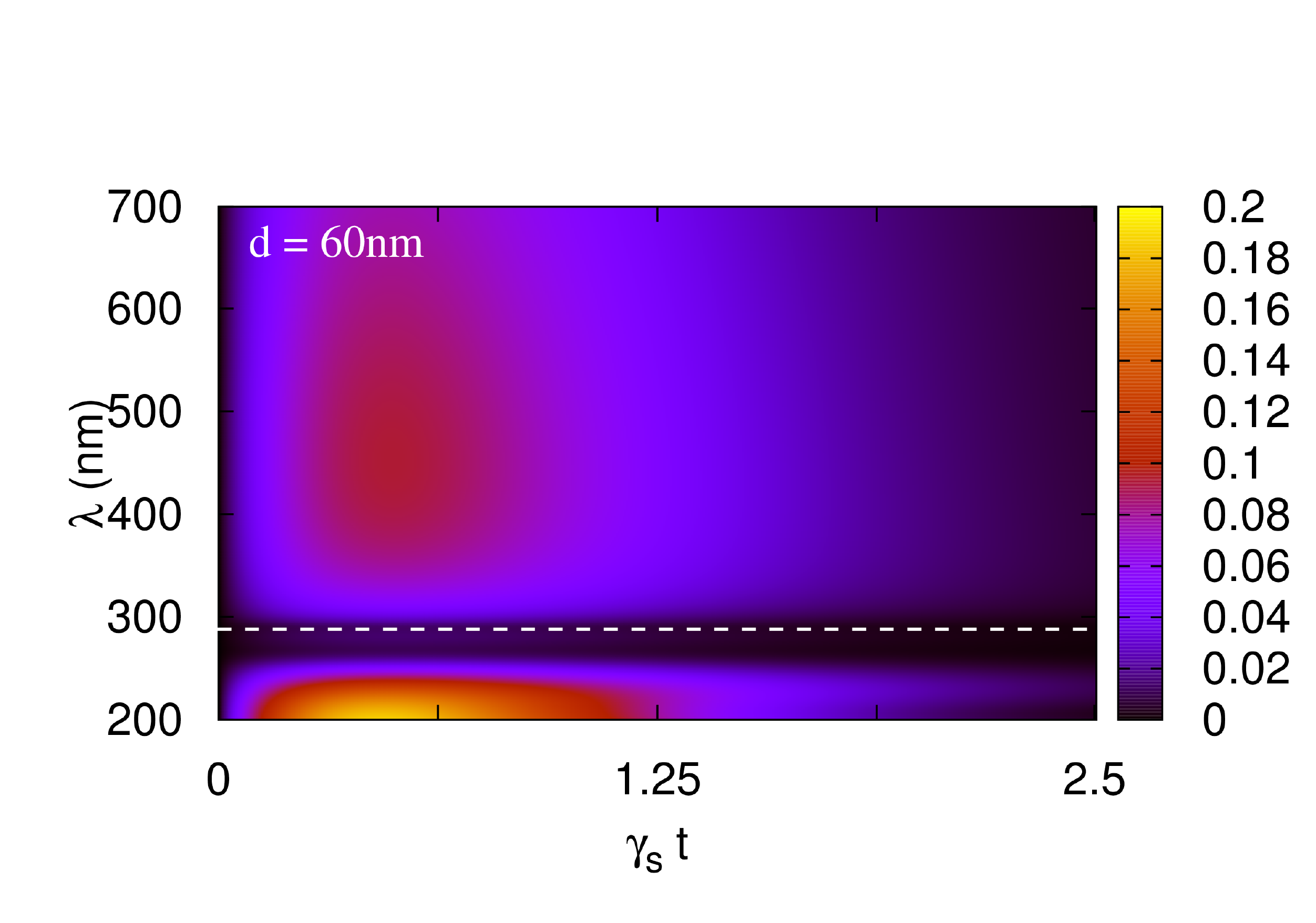} 
 \includegraphics[width = 0.45\textwidth]{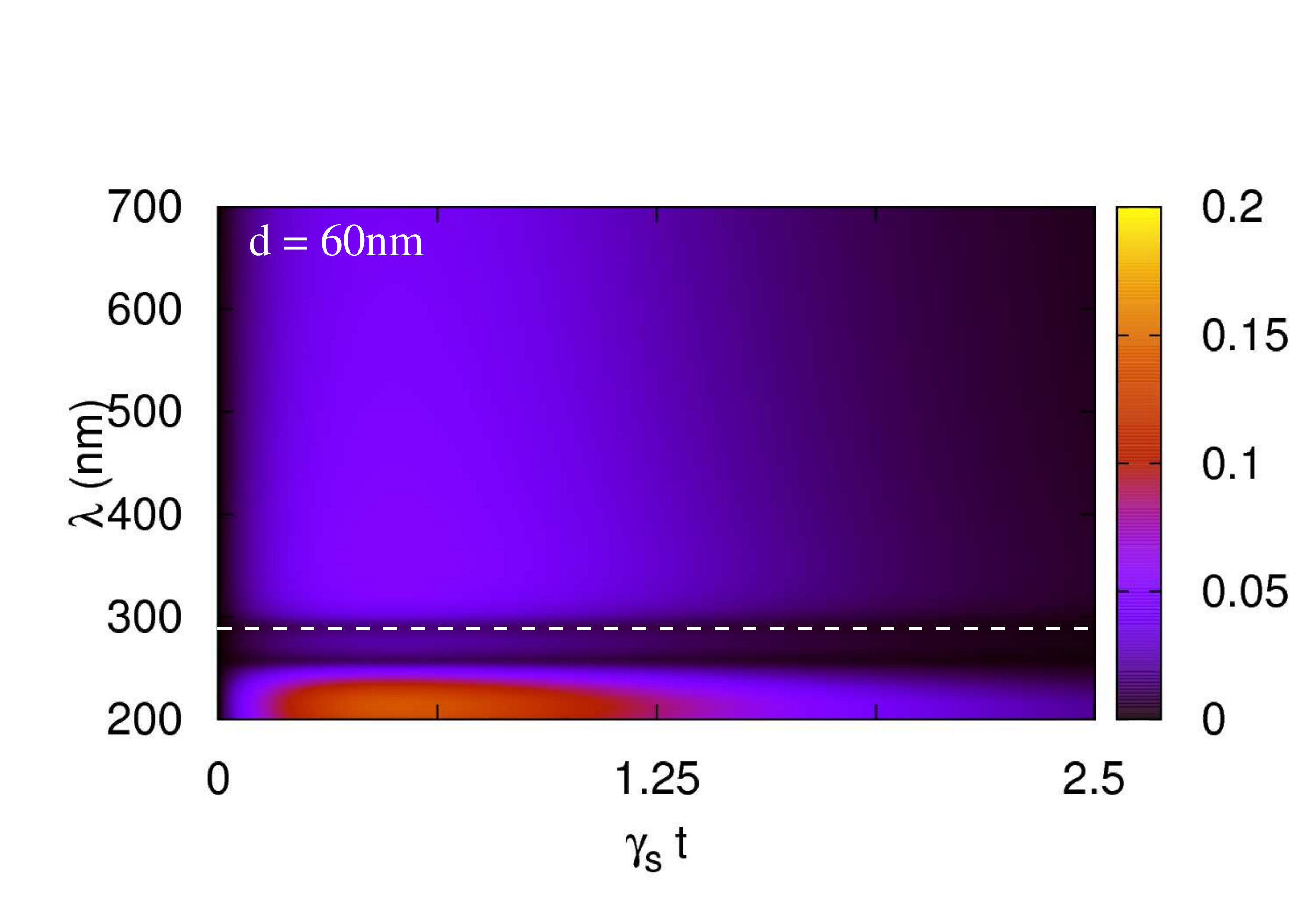} 
  \end{center}
  \caption{Concurrence function $C(t)$ for two TLS separated by a silver film of thicknesses $d = 10\,{\rm nm}$ and $60\,{\rm nm}$. Left column is for x-orientation of the dipole moments and right column for the z-orientation. The dashed horizontal line marks the surface plasmon wavelength $\lambda_{\rm SP} = 291\,{\rm nm}$.
\label{Fig:Silverfilm}}
\end{figure*}

\begin{figure*}[hbt]
  \begin{center}
  \includegraphics[width = 0.45\textwidth]{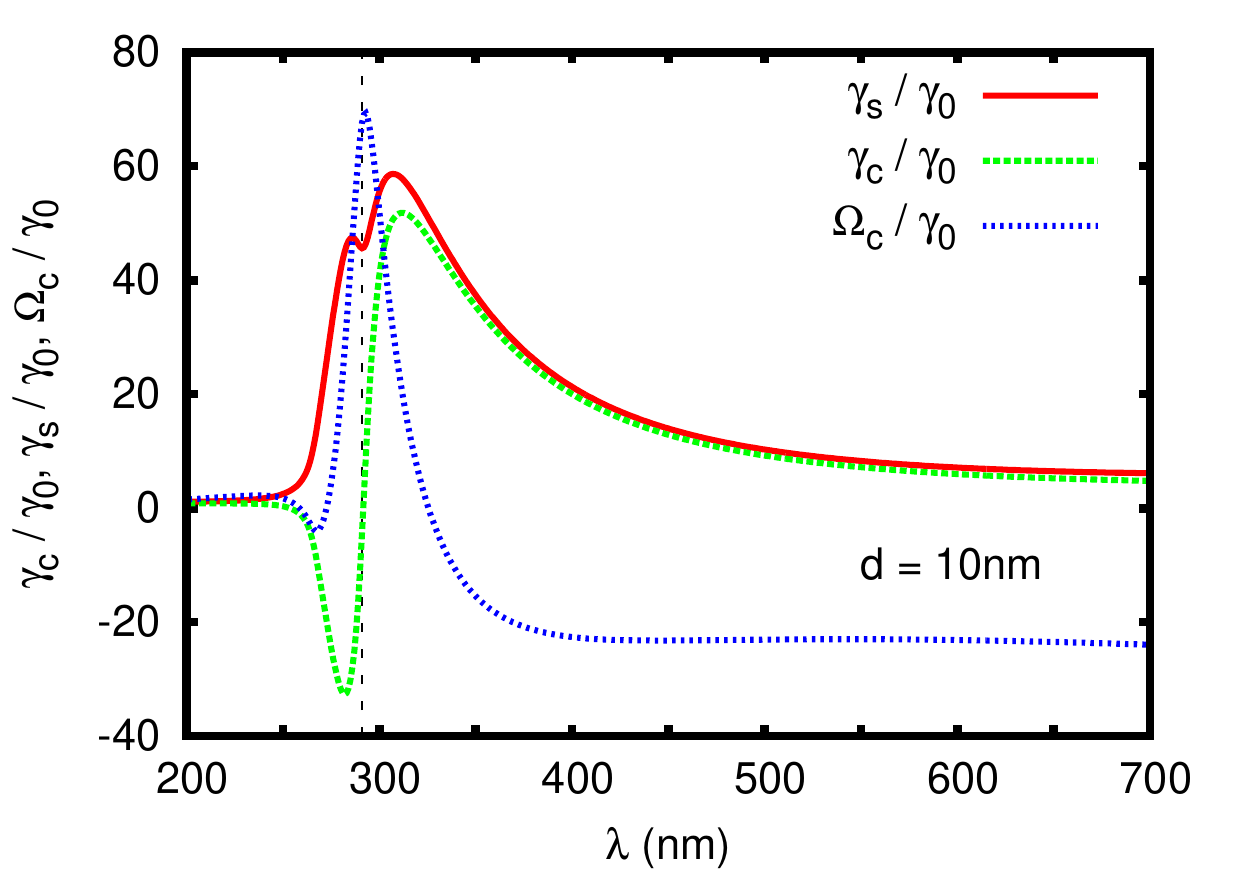} 
  \includegraphics[width = 0.45\textwidth]{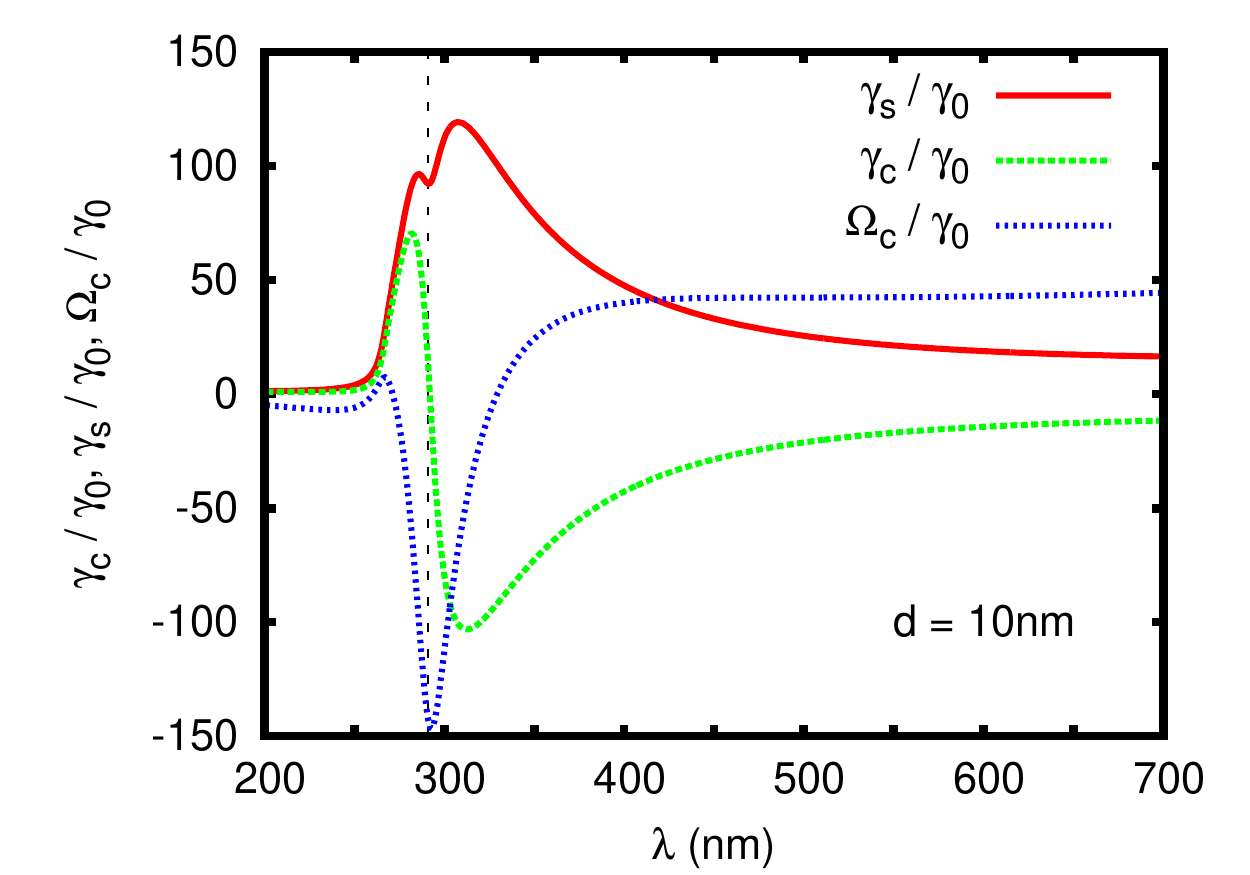} \\
  \includegraphics[width = 0.45\textwidth]{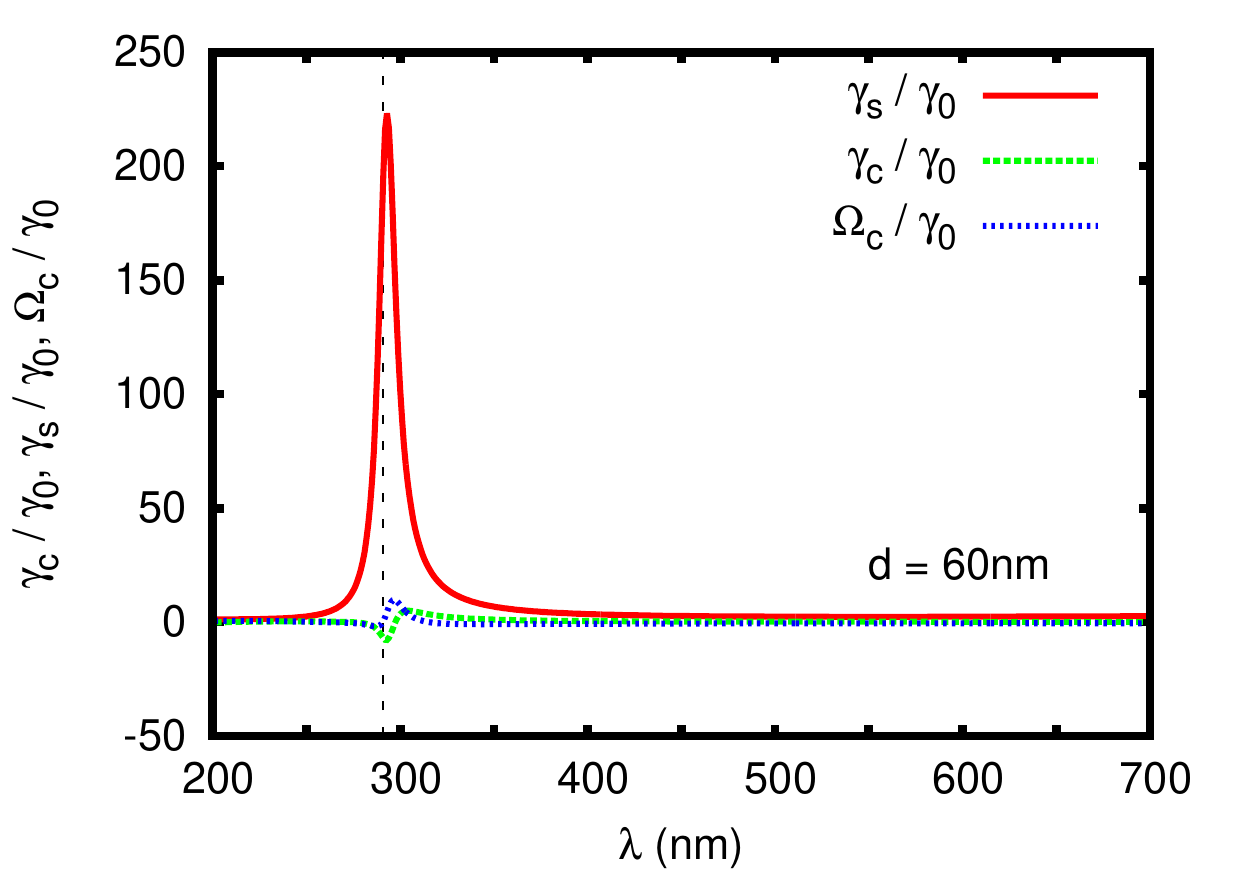}
  \includegraphics[width = 0.45\textwidth]{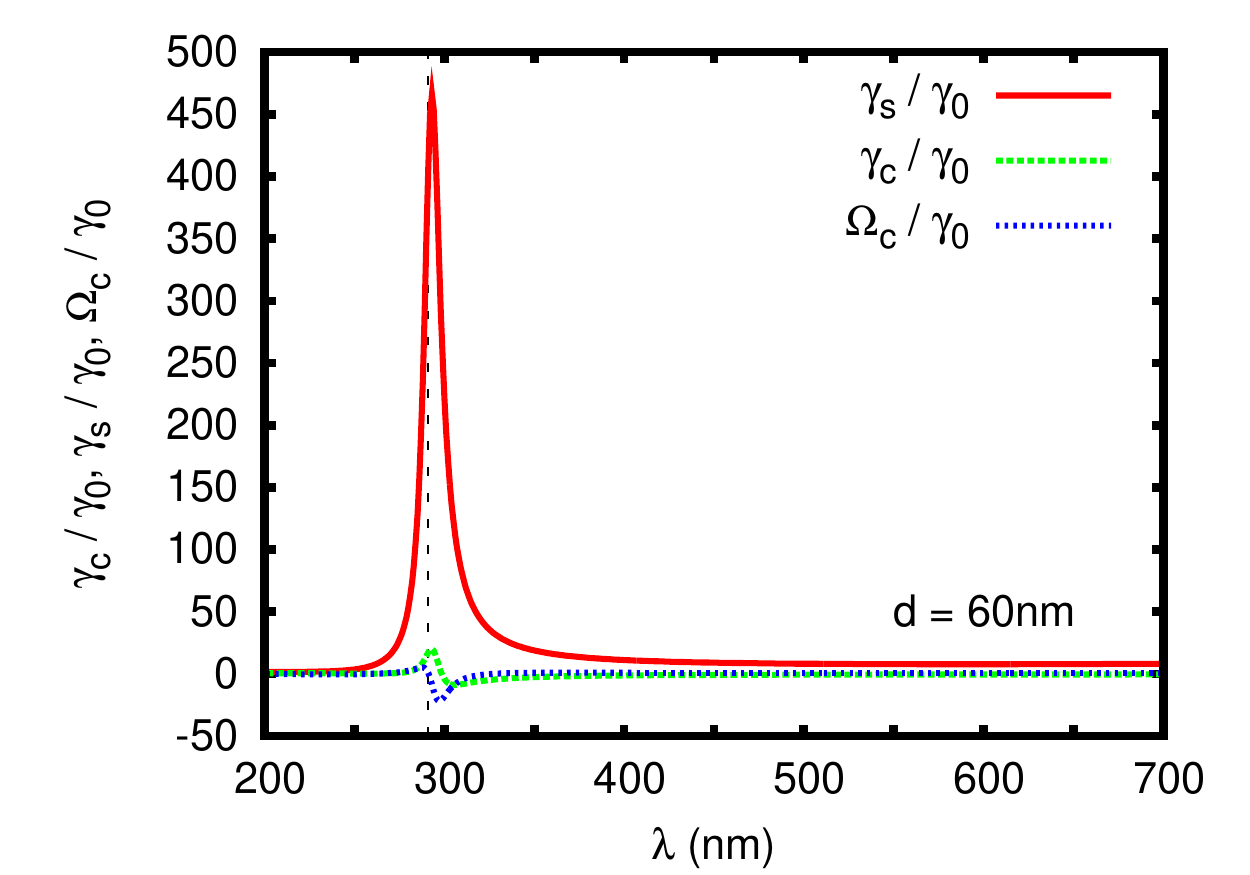}
  \end{center}
  \caption{$\gamma_\rc$, $\gamma_\rs$ and $\Omega_\rc$ for two TLS separated by a silver film of thicknesses $d = 10\,{\rm nm}$ and $60\,{\rm nm}$ normalized to the free space emission rate $\gamma_0$. Left column is for x-orientation of the dipole moments and right column for the z-orientation. The dashed vertical line marks the surface plasmon wavelength $\lambda_{\rm SP} = 291\,{\rm nm}$.
\label{Fig:SilverfilmABC}}
\end{figure*}

In Fig.~\ref{Fig:Silverfilm} we show some examples for the concurrence function $C(t)$ for 
silver films of different thickness. Throughout the paper we choose $z_1 = 10\,{\rm nm}$. 
It can be seen that for very thin films with thickness $d = 10\,{\rm nm}$
one can find a relatively large entanglement for $\lambda > \lambda_{\rm SP}$ of the two TLS due 
to the coupling via the coupled surface plasmons inside the metal film. On the other hand for thicker films 
with $d = 60\,{\rm nm}$ the entanglement for $\lambda > \lambda_{\rm SP}$ becomes already very
small and for $d = 120\,{\rm nm}$ it is practically not existing. To get more inside why this
happens we have plotted $\gamma_\rc$, $\gamma_\rs$ and $\Omega_\rc$ in Fig.~\ref{Fig:SilverfilmABC}.
As is clear from the expression for the concurrence function in Eq.~(\ref{Eq:Concurrence}) 
for $|\gamma_\rc| t \gg 1$ or $\Omega_\rc t \ll 1$ we have
\begin{equation}
  C(t) \approx \frac{1}{2}\re^{2 (|\gamma_\rc| -\gamma_\rs) t}. 
\end{equation}
Obviously, in this limit the concurrence can only have a maximum value of 0.5 and 
there can only be a noticable entanglement if the collective damping rate $|\gamma_\rc|$ 
is on the same order as the single-atom emission rate $\gamma_\rs$. Since the energy transmission 
rate is proportional to $|\mathds{G}_{xx/zz}(\mathbf{r_1,r_2})|^2 = \gamma_\rc^2 + \Omega_\rc^2$ this 
means that we can find a noticable entanglement if loosely speaking the 
'energy transmission rate' $|\gamma_\rc|$ is on the same order of magnitude as the 
single-atom spontaneous emission rate $\gamma_\rs$ of the initially excited TLS.

From Fig.~\ref{Fig:SilverfilmABC} it becomes apparent that the spontaneous emimssion rate $\gamma_\rs$
is very large around $\lambda_{\rm SP}$ as expected. Recently such changes in the spontaneous emission
have also been studied for three-level systems on meta-surfaces and hyperbolic materials~\cite{JhaEtAl2015,SunJiang2016}. In particular, the effect of the coupling of
the surface plasmons in the thin silver film $d = 10\,{\rm nm}$ can be nicely seen. The energy transmission
rate is also very large for wavelengths around $\lambda_{\rm SP}$ as expected, but it drops rapidely
when the film thickness is increased. This is so because the coupling between the surface plasmons on both
interfaces becomes very small when $d$ is increased due to the evanescent nature of the surface plasmon polariton
modes. This leads to a less efficient coupling of both TLS and therefore to very small entanglement for
thicker metal films. {Note, that at the ENZ wavelength of 259nm there is no significant
effect of increased entanglement. The relatively large entanglement which can be seen in Fig.~\ref{Fig:Silverfilm} 
for thick films is in the transparency region of the silver film for $\lambda < 259\,{\rm nm}$. }

In order to contrast the results obtained for metal films, we consider as a second structure 
a multilayer hyperbolic meta-material of alternating Ag and TiO$_2$ layers. The effective 
permittivities are for this structure given by
\begin{align}
  \epsilon_\perp &= f \epsilon_{\rm Ag} + (1 - f) \epsilon_{{\rm TiO}_2}, \\
  \epsilon_\parallel     &= \frac{\epsilon_{\rm Ag} \epsilon_{{\rm TiO}_2}}{f \epsilon_{{\rm TiO}_2} + (1-f) \epsilon_{\rm Ag} },
\end{align}
where $f$ is the filling fraction of silver and $\epsilon_{\rm Ag}$/$\epsilon_{{\rm TiO}_2}$
are the permittivites of the both constitutents of the multilayer structure. For silver we use the Drude 
model in Eq.~(\ref{Eq:Drude}). TiO$_2$ is transparent in the visible regime. It's 
permittivity $\epsilon_{{\rm TiO}_2}$ is nearly constant in that regime and can be well described 
by the formula~\cite{Devore1951}
\begin{equation}
  \epsilon_{{\rm TiO}_2} = 5.913 + \frac{0.2441}{\lambda^2 - 0.0803}.
\end{equation}
As shown in Ref.~\cite{BiehsEtAl2016} when choosing $f = 0.35$ this multilayer structure has a type I hyperbolic band [$\Re(\epsilon_\parallel) < 0$ and $\Re(\epsilon_\perp) > 0$] at wavelengths below the epsilon-near-pole wavelength $\lambda_{\rm ENP} = 395\,{\rm nm}$ and a type II hyperbolic band [$\Re(\epsilon_\parallel) > 0$ and $\Re(\epsilon_\perp) < 0$] above the epsilon-near-zero wavelength $\lambda_{\rm ENZ} = 551\,{\rm nm}$. For
$\lambda_{\rm ENP} < \lambda < \lambda_{\rm ENZ}$ the multilayer structure behaves like a normal uni-axial
dielectric [$\Re(\epsilon_\parallel) > 0$ and $\Re(\epsilon_\perp) > 0$]. 

\begin{figure*}[hbt]
  \begin{center}
   \includegraphics[width = 0.45\textwidth]{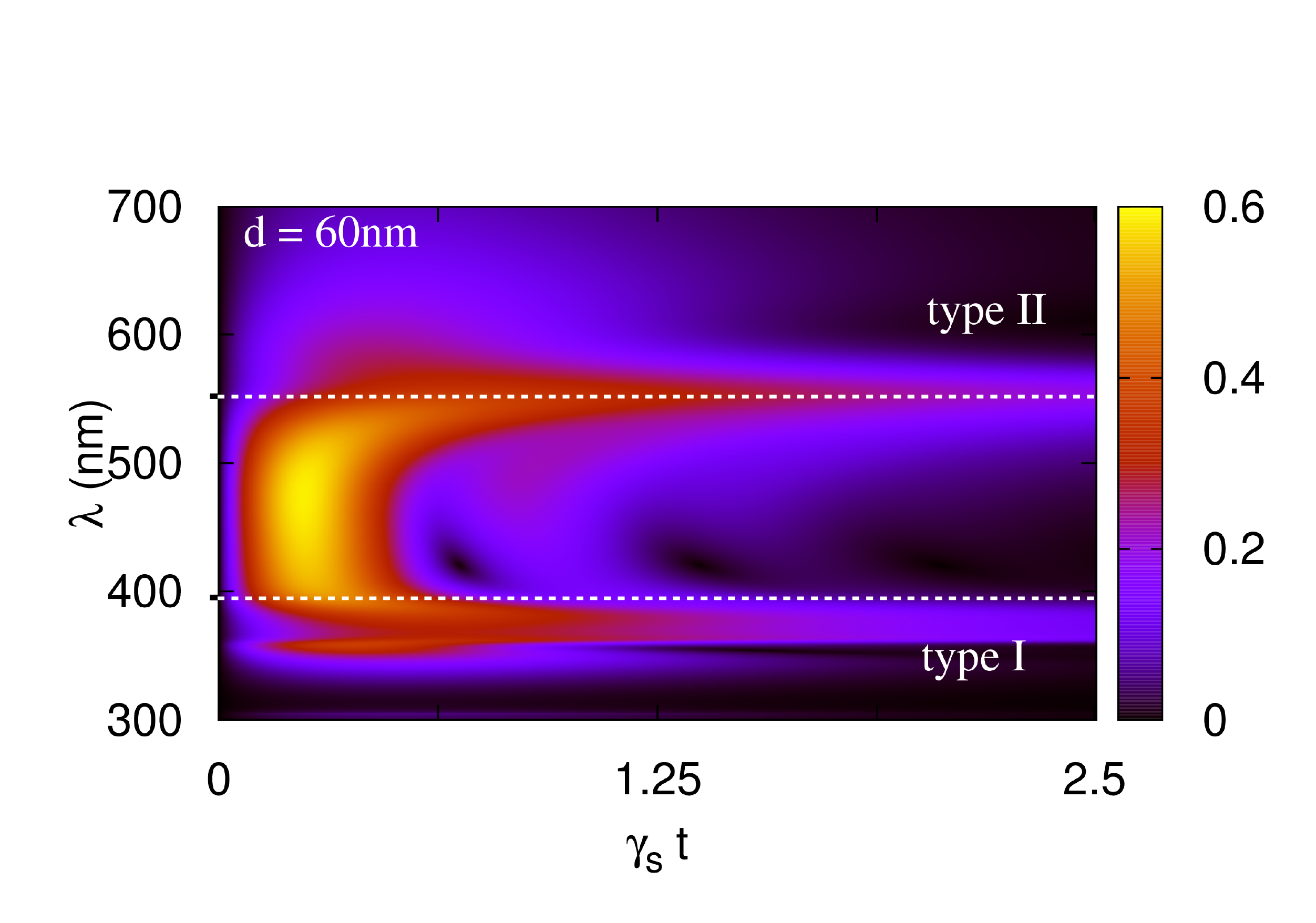}
   \includegraphics[width = 0.45\textwidth]{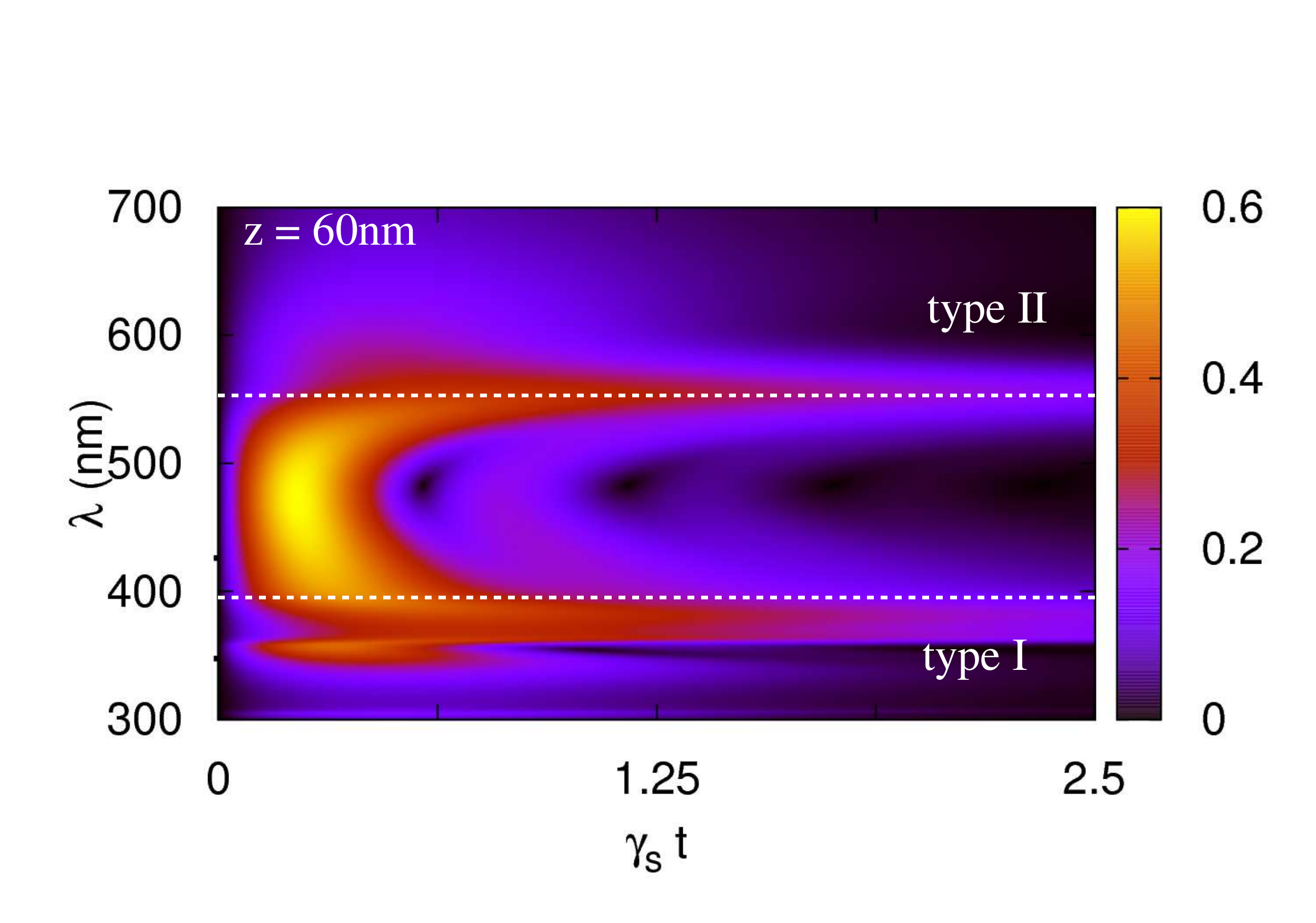} \\
   \includegraphics[width = 0.45\textwidth]{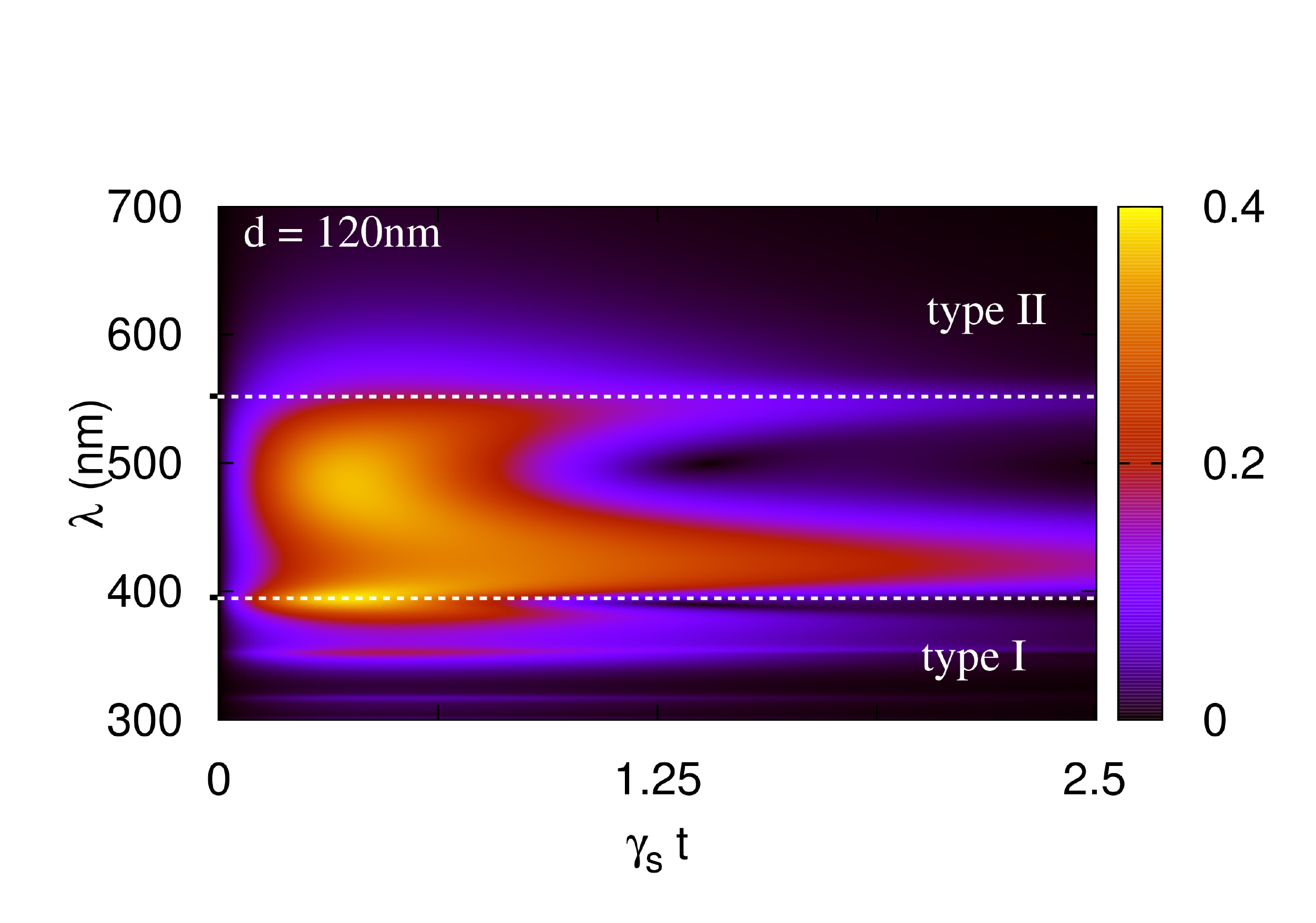}
   \includegraphics[width = 0.45\textwidth]{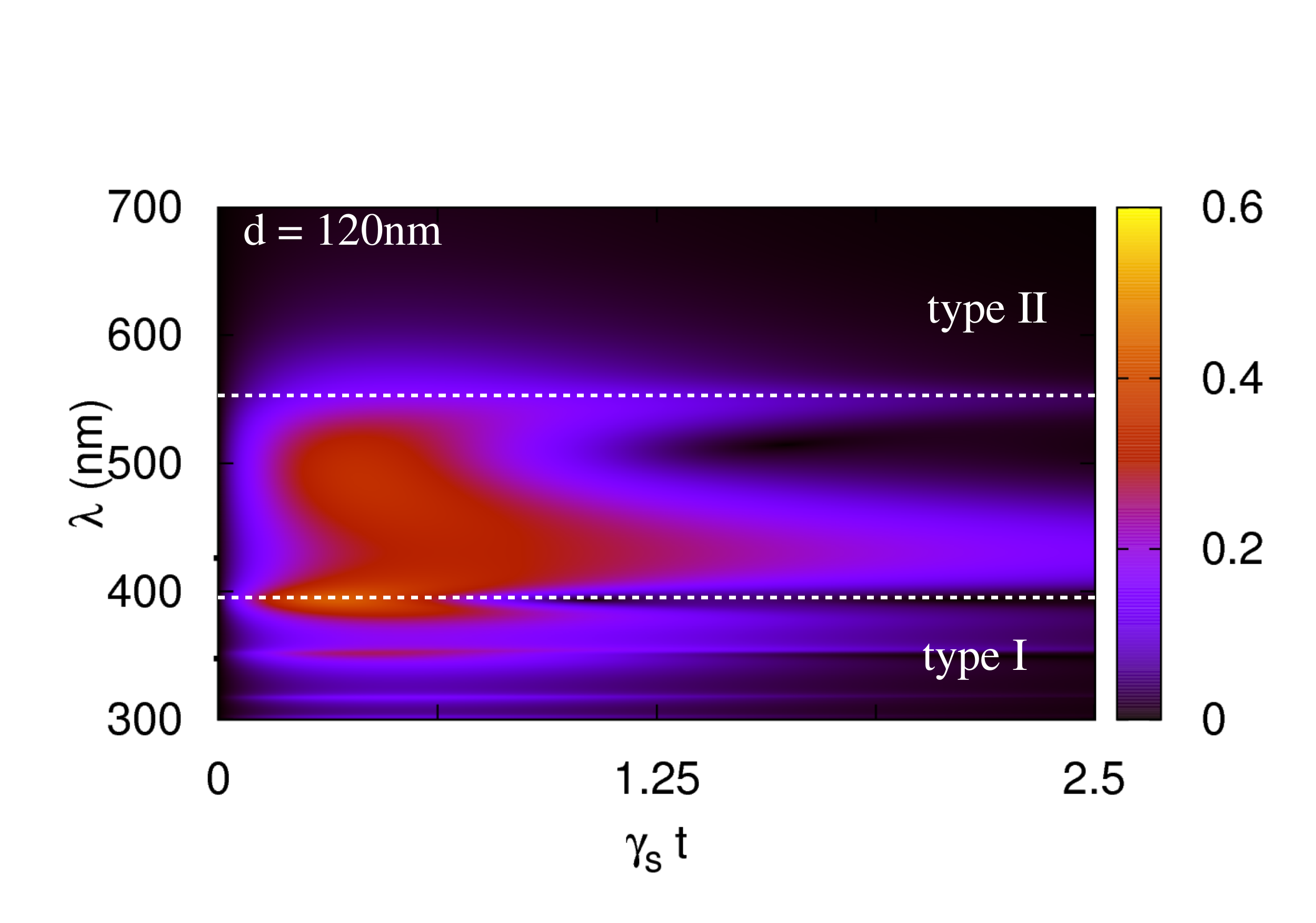}
  \end{center}
  \caption{Concurrence function $C(t)$ for two TLS separated by a HMM film of thicknesses $d = 60\,{\rm nm}$ and $120\,{\rm nm}$. Left column is for x-orientation of the dipole moments and right column for the z-orientation. The horizontal dashed lines mark the ENP and ENZ wavelengths $\lambda_{\rm ENP} = 395\,{\rm nm}$ and $\lambda_{\rm ENZ} = 551\,{\rm nm}$.
\label{Fig:HMMfilm}}
\end{figure*}

\begin{figure*}[hbt]
  \begin{center}
   \includegraphics[width = 0.45\textwidth]{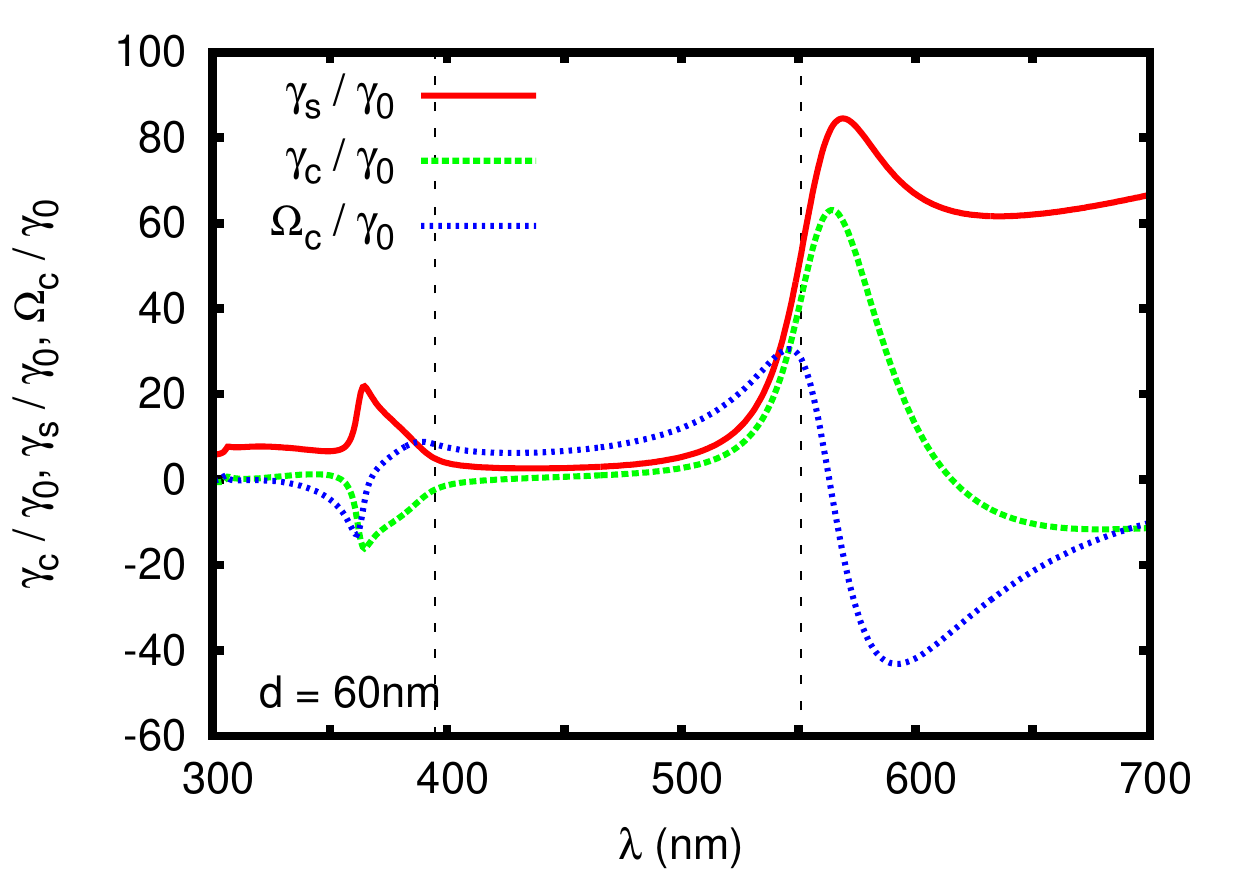}
   \includegraphics[width = 0.45\textwidth]{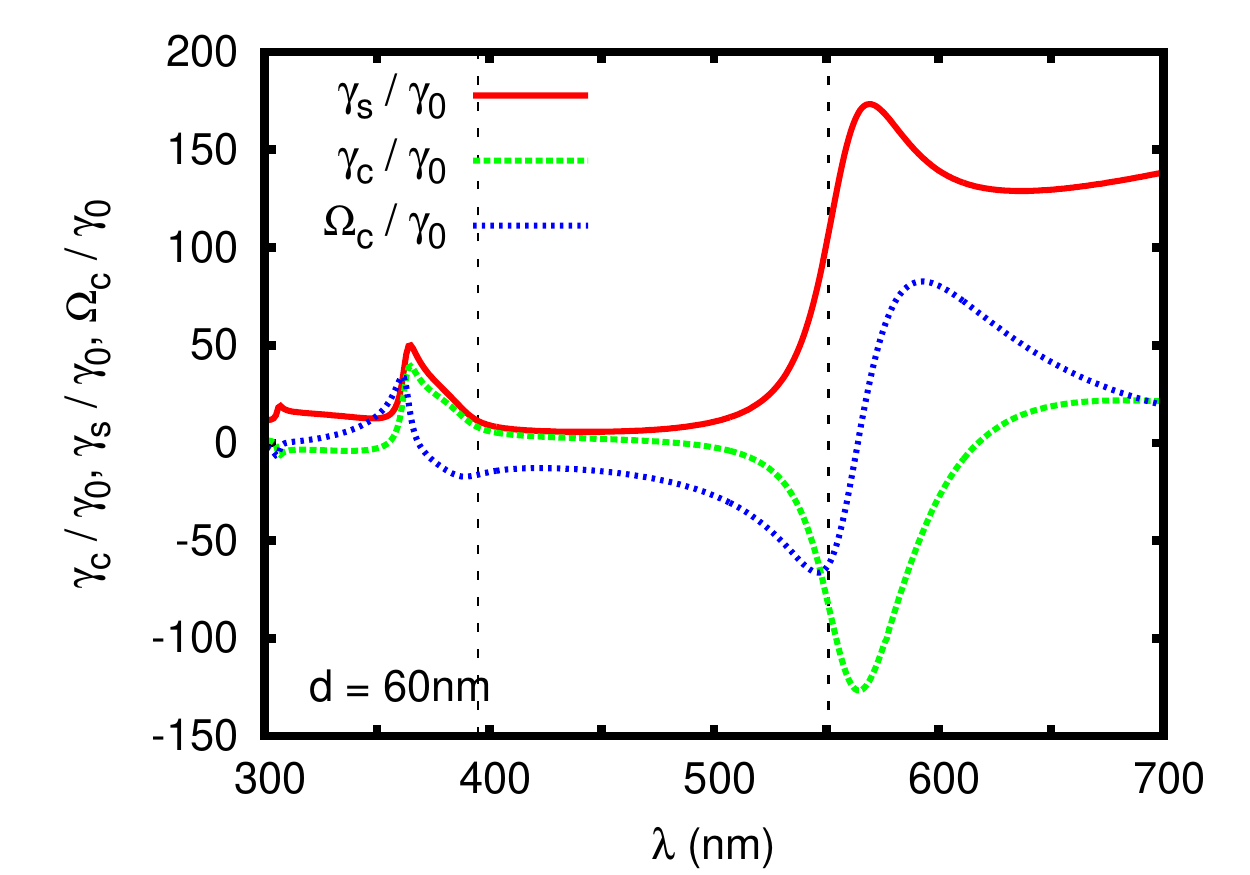}\\
   \includegraphics[width = 0.45\textwidth]{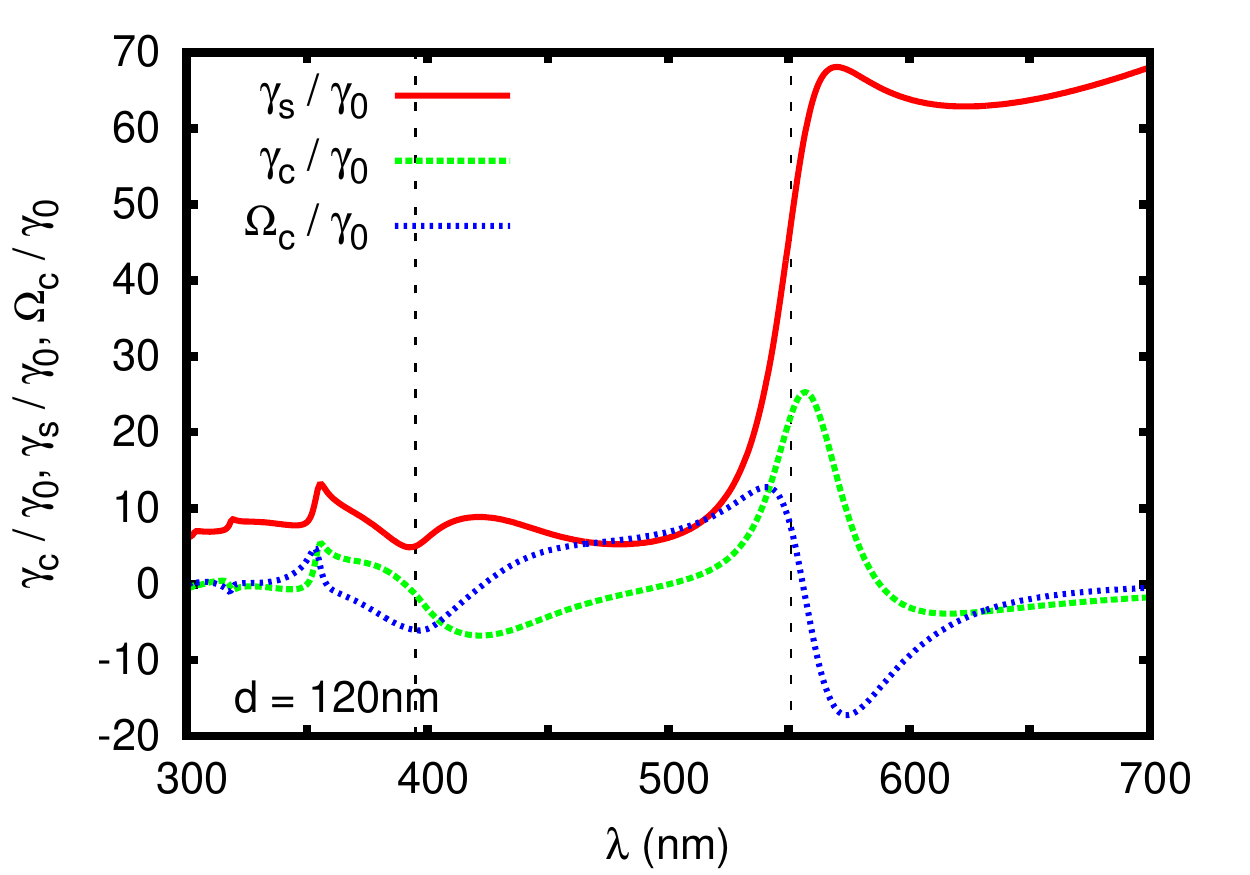}
   \includegraphics[width = 0.45\textwidth]{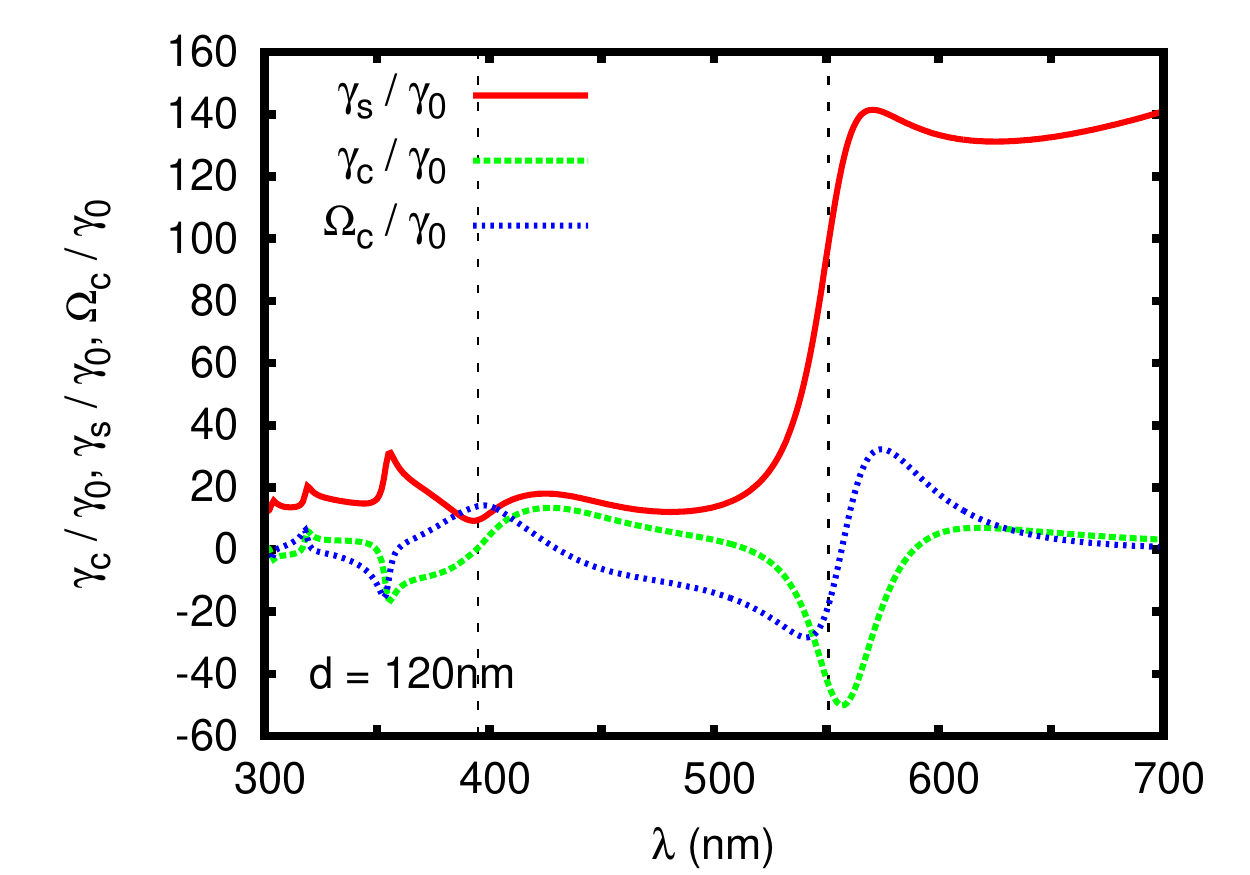}
  \end{center}
 \caption{$\gamma_\rc$, $\gamma_\rs$ and $\Omega_\rc$ for two TLS separated by a HMM film of thicknesses $d = 60\,{\rm nm}$ and $120\,{\rm nm}$ normalized to the free space emission rate $\gamma_0$. Left column is for x-orientation of the dipole moments and right column for the z-orientation. The vertical dashed lines mark the ENP and ENZ wavelengths $\lambda_{\rm ENP} = 395\,{\rm nm}$ and $\lambda_{\rm ENZ} = 551\,{\rm nm}$.
\label{Fig:HMMfilmABC}}
\end{figure*}

In Fig.~\ref{Fig:HMMfilm} we show the concurrence function for the HMM as a function of time and
wavelength for $d = 60\,{\rm nm}$ and $120\,{\rm nm}$. It can be seen 
that for $d = 60\,{\rm nm}$ the entanglement is especially large at the ENZ and ENP wavelength. 
For $d = 120\,{\rm}$ we have still a relatively large entanglement 
especially close to the ENZ wavelength and in the normal dielectric region with $\lambda_{\rm ENP} < \lambda < \lambda_{\rm ENZ}$. As can be seen in Fig.~\ref{Fig:HMMfilmABC} this is in agreement
with the large transmission at the ENZ wavelength which has been studied in very much detail 
in Ref.~\cite{BiehsEtAl2016}. The difference to the silver film is that
here we have propagating modes inside the hyperbolic material with large wavevectors allowing a
strong coupling between the two TLS even for relatively thick films. The main limiting factor
for this strong coupling is the damping of these propagating modes inside the hyperbolic 
materials. Apart from the fact that we can have large entanglement for relatively thick films 
hyperbolic materials have the advantage that the position of the ENP and ENZ wavelength can 
be engineered by the combination of different materials and by changing the filling fraction of the metal part
of the structure so that the wavelength for which a strong coupling is needed can be adapted
at will.

%
%

\section{Conclusion}

\begin{figure}
  \includegraphics[width = 0.45\textwidth]{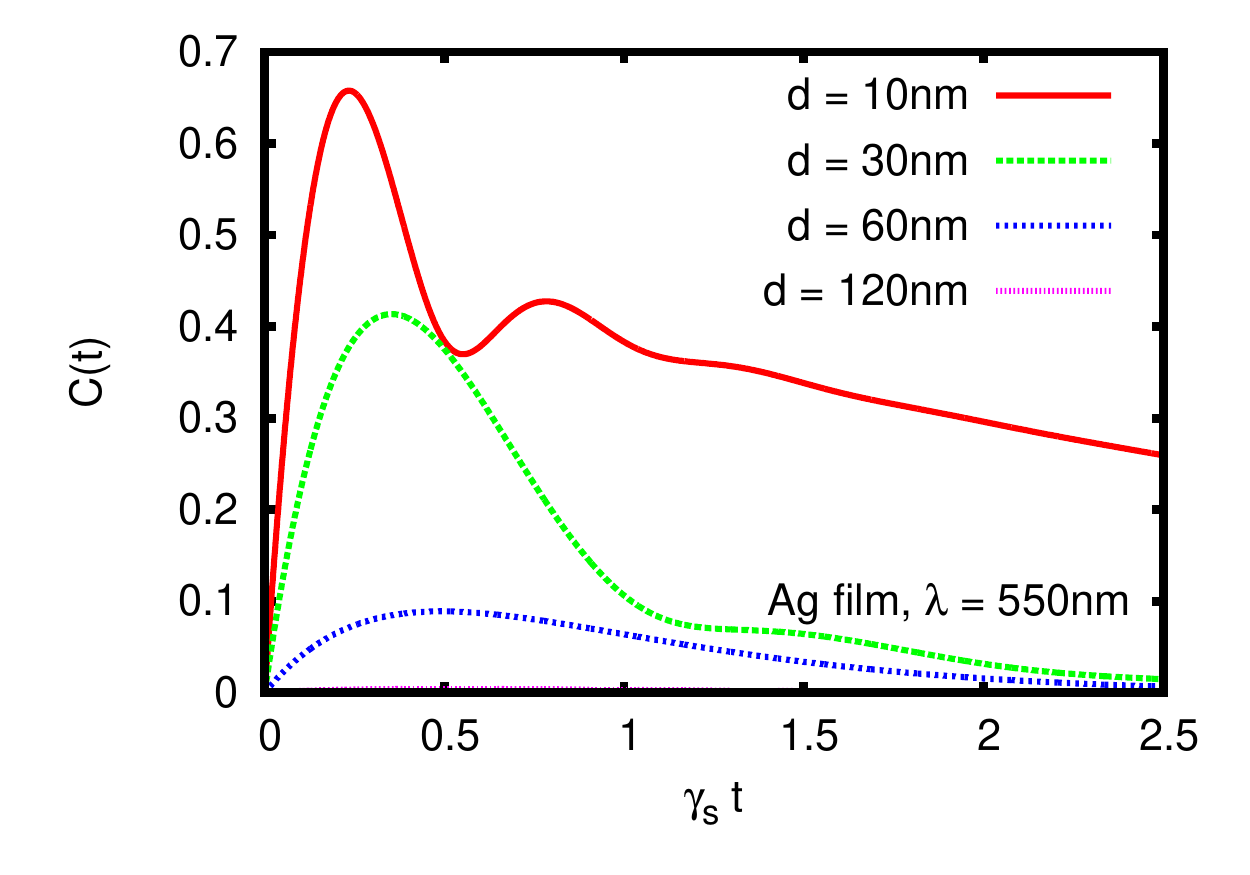}
  \includegraphics[width = 0.45\textwidth]{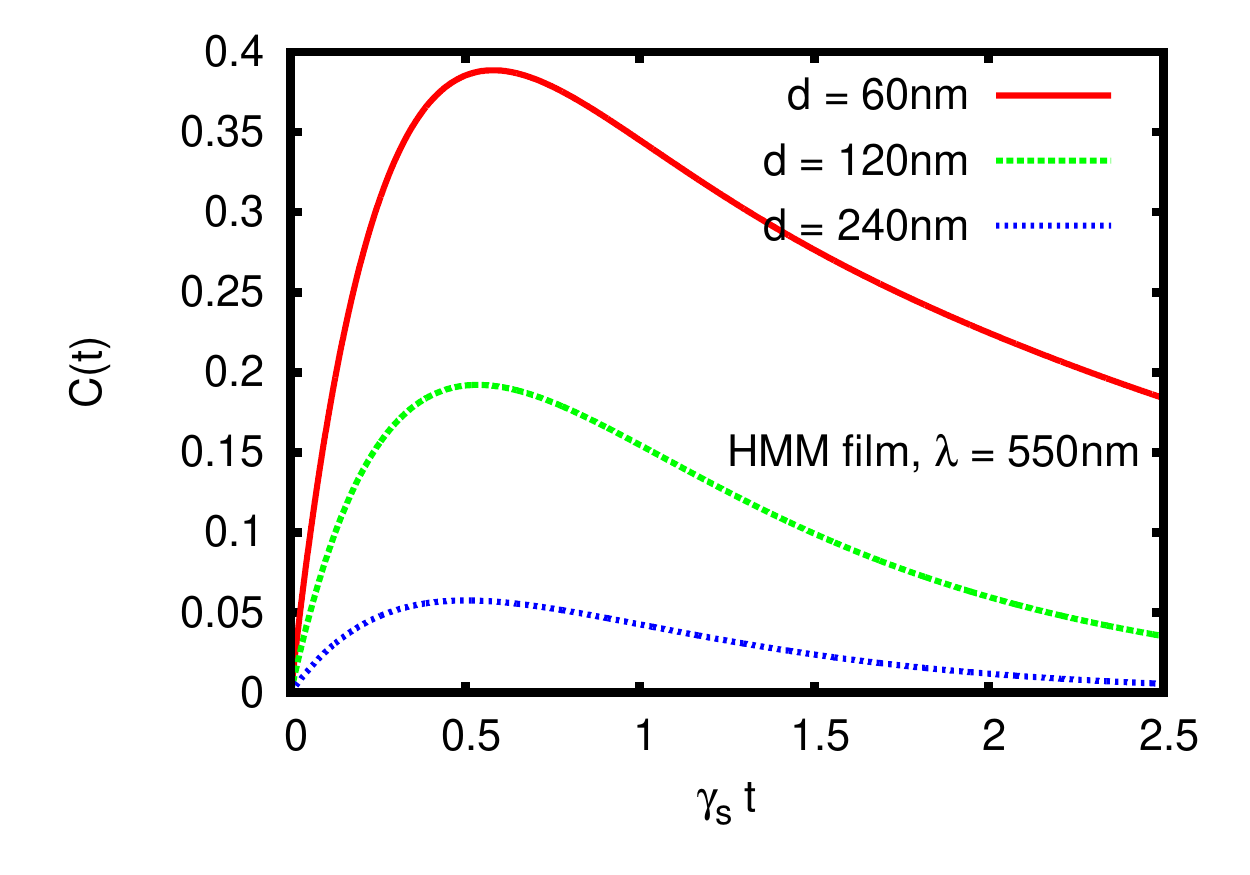}
  \caption{Concurrence function for two TLS separated by an Ag and a HMM film with x-oriented dipole moments for different thicknesses evaluated at the ENZ wavelength $\lambda = 550\,{\rm nm}$ of the HMM.} 
\label{Fig:VacHMMfilmABC}
\end{figure}

To summarize, we have studied the entanglement of two TLS separated by a thin film
using the master-equation approach and the concurrence function as entanglement measure.
We have compared the entanglement as function of the film thickness of the intermediate
layer for a silver film and a multilayer Ag/TiO$_2$ hyperbolic meta-material. Our main
finding is summarized in Fig.~\ref{Fig:VacHMMfilmABC} where the concurrence function is
plotted for the silver film and the hyperbolic material close to the ENZ wavelength
at 550nm for different film thicknesses. At this wavelength the single-atom emission 
rates and the energy transmission rates for the TLS in presence of the hyperbolic material
 are very large compared to the vacuum value. By changing the filling fraction of the
silver layers in the Ag/TiO$_2$ hyperbolic meta-material one can shift this important
frequency. It can be seen in Fig.~\ref{Fig:VacHMMfilmABC} that for the hyperbolic material 
one can find the same entanglement as for the metallic film but for twice as thick layers.
At the ENZ wavelength of the metallic film we find in most
cases a relatively small entanglement as is shown in Fig.~\ref{Fig:VacHMMfilmABC} so that 
for operation at the ENZ wavelengths hyperbolic metamaterial are much more advantageous 
than metal films. We believe that by optimizing the vertical and horizontal positions of 
the TLS and by optimizing the multilayer layout one can achieve substantial entanglement 
even for thicker films.

%
%

\section*{acknowledgments}

The authors thank the Bio Photonics initiative of the Texas A \& M university for supporting this work.
S.A.B.~thanks the hospitality of the Texas A \& M university.

\appendix

\section*{Appendix}

\section{Derivation of $\mathds{G}_{xx}$ and $\mathds{G}_{zz}$}

If the dipole moments of the TLS are oriented both in $z$ or $x$ directions. In this case the 
above expressions can be further simplified.

\subsection{Dipoles in z-direction}

If the dipole moments are oriented in z-direction we find
\begin{align}
  [\mathbf{a}_\rs^\pm(k_0) \otimes \mathbf{a}_\rs^\pm(k_0)]_{zz} &= 0, \\
  [\mathbf{a}_\rp^\pm(k_0) \otimes \mathbf{a}_\rp^\pm(k_0)]_{zz} &= \frac{\kappa^2}{k_0^2}
\end{align}
and therefore
\begin{equation}
   \mathds{G}_{zz} (\mathbf{r}_1, \mathbf{r}_2) = \int \frac{\rd \kappa}{2 \pi} \, \kappa \frac{\ri \re^{\ri k_{z,{\rm vac}} (d + 2|z_1|)}}{2 k_{z,{\rm vac}}} t_\rp \frac{\kappa^2}{k_0^2}
\end{equation}
and
\begin{equation}
  \mathds{G}_{zz}^{\rm single} (\mathbf{r}_1, \mathbf{r}_1) = \int \frac{\rd \kappa}{2 \pi} \, \kappa \frac{\ri}{2 k_{z,{\rm vac}}} \frac{\kappa^2}{k_0^2} \biggl(1 + r_\rp \re^{2 \ri k_{z,{\rm vac}} |z_1|} \biggr).
\end{equation}
From this expression we can retrieve the results for the case where the TLS are only coupled
by vacuum by setting $t_\rp = 1$ and $r_\rp = 0$ so that
\begin{equation}
   \mathds{G}_{zz}^{\rm vac} (\mathbf{r}_1, \mathbf{r}_2) = \int \frac{\rd \kappa}{2 \pi} \, \kappa \frac{\ri \re^{\ri k_{z,{\rm vac}} (d + 2|z_1|)}}{2 k_{z,{\rm vac}}} \frac{\kappa^2}{k_0^2}
\end{equation}
and
\begin{equation}
  \mathds{G}_{zz}^{\rm single, vac} (\mathbf{r}_1, \mathbf{r}_1)  = \int \frac{\rd \kappa}{2 \pi} \, \kappa \frac{\ri}{2 k_{z,{\rm vac}}} \frac{\kappa^2}{k_0^2}.  
\end{equation}
The emission rate for the single atom in vacuum is in this case
\begin{equation}
  \gamma_\rs^{\rm vac} = \frac{6 \pi \gamma_0}{k_0} \Im (\mathds{G}_{zz}) =  \frac{6 \pi \gamma_0}{k_0} \frac{1}{4 \pi} k_0 \frac{2}{3} = \gamma_0.
\end{equation}

\subsection{Dipoles in x-direction}

If the dipole moments of the TLS are oriented in x-direction we find
\begin{align}
  [\mathbf{a}_\rs^\pm(k_0) \otimes \mathbf{a}_\rs^\pm(k_0)]_{xx} &= \frac{k_y^2}{\kappa^2}, \\
  [\mathbf{a}_\rp^\pm(k_0) \otimes \mathbf{a}_\rp^\pm(k_0)]_{xx} &= \frac{k_x^2 k_{z,{\rm vac}}^2}{\kappa^2 k_0^2}.
\end{align}
Introducing polar coordinates for $\boldsymbol{\kappa} = \kappa (\cos \theta, \sin \theta)^t$ we find therefore
\begin{equation}
   \mathds{G}_{xx} (\mathbf{r}_1, \mathbf{r}_2) = \int \frac{\rd \kappa}{2 \pi} \, \kappa \frac{\ri \re^{\ri k_{z,{\rm vac}} (d + 2|z_1|)}}{2 k_{z,{\rm vac}}} \frac{1}{2} \biggl( t_\rs + t_\rp \frac{k_{z,{\rm vac}}^2}{k_0^2} \biggr)
\end{equation}
and
\begin{equation}
\begin{split}
  \mathds{G}_{xx}^{\rm single} (\mathbf{r}_1, \mathbf{r}_1) &= \int \frac{\rd \kappa}{2 \pi} \, \kappa \frac{\ri}{2 k_{z,{\rm vac}}} \biggl[\frac{k_0^2 + k_{z,{\rm vac}}^2}{2 k_0^2} \\
              &\qquad+ \re^{2 \ri k_{z,{\rm vac}} |z_1|} \frac{1}{2} \biggl( r_\rs - r_\rp \frac{k_{z,{\rm vac}}^2}{k_0^2} \biggr) \biggr].
\end{split}
\end{equation}
Again we can retrieve the relations for the case where both TLS are coupled by vacuum by setting $t_\rp = 1$ and $r_\rp = 0$ so that
\begin{equation}
   \mathds{G}_{xx}^{\rm vac} (\mathbf{r}_1, \mathbf{r}_2) = \int \frac{\rd \kappa}{2 \pi} \, \kappa \frac{\ri \re^{\ri k_{z,{\rm vac}} (d + 2|z_1|)}}{2 k_{z,{\rm vac}}} \frac{1}{2} \biggl( 1 + \frac{k_{z,{\rm vac}}^2}{k_0^2} \biggr)
\end{equation}
and
\begin{equation}
  \mathds{G}_{xx}^{\rm single, vac} (\mathbf{r}_1, \mathbf{r}_1)  =  \int \frac{\rd \kappa}{2 \pi} \, \kappa \frac{\ri}{2 k_{z,{\rm vac}}} \frac{k_0^2 + k_{z,{\rm vac}}^2}{2 k_0^2} 
\end{equation}
The emission rate for the single atom in vacuum is in this case again
\begin{equation}
  \gamma_\rs^{\rm vac} = \frac{6 \pi \gamma_0}{k_0} \Im (\mathds{G}_{xx}) =  \frac{6 \pi \gamma_0}{k_0} \frac{1}{4 \pi} k_0 \frac{2}{3} = \gamma_0.
\end{equation}


\end{document}